\documentclass[]{IEEEtran}
\IEEEoverridecommandlockouts
\usepackage[T1]{fontenc}

\usepackage{newtxtext}
\usepackage[varvw]{newtxmath}
\usepackage{tikz}
\usepackage{url}

\usetikzlibrary{patterns}
\usetikzlibrary{arrows}

\tikzset{
     task7/.style={draw,minimum height=\uy,},
}
\def\ux{0.35cm}
\def\uy{0.5cm} 

\usepackage{graphicx}

\newtheorem{theorem}{Theorem}
\newtheorem{definition}{Definition}

\begin{document}
 
\title{\LARGE Comments Paper: Counterexample for Schedulability Analysis for
  Self-Suspending Tasks under EDF-Like Scheduling}
\author{
  \IEEEauthorblockN{Michael Frichert\IEEEauthorrefmark{1}, Mario Günzel\IEEEauthorrefmark{2}, Marion Sudvarg\IEEEauthorrefmark{3}, and Jian-Jia Chen\IEEEauthorrefmark{1}}\\
  \IEEEauthorblockA{\IEEEauthorrefmark{1}RWTH Aachen University, \IEEEauthorrefmark{2}MPI-SWS, \IEEEauthorrefmark{3}Missouri University of Science and Technology\\
    \IEEEauthorrefmark{1}\{michael.frichert, jian-jia.chen\}@rwth-aachen.de, \IEEEauthorrefmark{2}mguenzel@mpi-sws.org, \IEEEauthorrefmark{3}msudvarg@mst.edu}
}
\maketitle              

\begin{abstract}
This paper provides a counterexample of the sufficient schedulability test
for self-suspending real-time tasks under EDF-Like scheduling given by Wang et al. in Theorem 11 of IEEE Trans.
Computers, 74(7):2364–2375, 2025. This paper also shows that their necessary condition is unsound.
\end{abstract}

\section{Introduction}

\emph{Self-suspension} occurs when a task pauses its execution voluntarily and
has been an active research area since 1988.  A 2019 review paper by
Chen et al.~\cite{suspension-review-jj} shows that several analytical results for self-suspending tasks have been flawed.

This paper provides a counterexample of the block-based schedulability
analysis in Theorem 11 by
Wang~et~al.~\cite{DBLP:journals/tc/WangLZLT25}, which disproves their
sufficient schedulability test. This paper also shows that the necessary
condition used in \cite{DBLP:journals/tc/WangLZLT25} is unsound.

We consider a set of constrained-deadline\footnote{Although the paper by Wang et al.~\cite{DBLP:journals/tc/WangLZLT25} even allows arbitrary deadline tasks, we show that the results are incorrect even for constrained deadline tasks.} sporadic real-time tasks
$\mathbb{T} = \{\tau_1, \dots, \tau_n\}$, in which each task $\tau_i = (C_i, S_i, D_i, T_i)$
is characterized by its minimum inter-arrival time $T_i$, relative deadline $D_i$, worst-case
self-suspension time $S_i$, and worst-case execution time $C_i$. The
relative deadline $D_i$ is set to be no more than
$T_i$. Wang~et~al.~\cite{DBLP:journals/tc/WangLZLT25} consider the
dynamic self-suspension model, in which a task instance (job) released
by a task $\tau_i$ can suspend arbitrarily as long as the total 
suspension time of the job does not exceed $S_i$.

Wang~et~al.~\cite{DBLP:journals/tc/WangLZLT25} explore EDF-like
scheduling \cite{DBLP:conf/rtss/GunzelBCC22}, in which the priority of
a job is defined by its arrival time plus its \emph{priority point}, denoted
$P_i$. This generalizes EDF, which is the special case of EDF-Like scheduling where
$P_i=D_i$.  For the rest of this paper, it is sufficient to provide
counterexamples using EDF, i.e., $P_i=D_i$ for all tasks
$\tau_i \in \mathbb{T}$.

Wang~et~al.~\cite{DBLP:journals/tc/WangLZLT25} state the following necessary condition
for schedulability of $\mathbb{T}$.

\begin{definition}[NecessaryConditions($\mathbb{T}$)~\cite{DBLP:journals/tc/WangLZLT25}]
  \label{def:necessary_cond}
  For each $\tau_i \in \mathbb{T} = \{\tau_1, \ldots, \tau_n\}$
  \begin{equation}
    \frac{S_i}{T_i} + \sum_{\tau_j \in \mathbb{T}} \frac{C_j}{T_j} \leq 1.
  \end{equation}
\end{definition}
  
\medskip

The analysis by Wang~et~al.~\cite{DBLP:journals/tc/WangLZLT25} for preemptive EDF scheduling can be rephrased as follows: 

\begin{theorem}[Theorem~11~in~\cite{DBLP:journals/tc/WangLZLT25}]
\label{th:sched_test}
Given a task set $\mathbb{T} = \{\tau_1, \ldots, \tau_n\}$ and $\mathbb{T}$ is sorted according to the relative deadlines in non-decreasing order. If the pre-condition in Definition~\ref{def:necessary_cond} holds, and for any $k = 1, \ldots, n$,
\begin{equation}
    \label{eq:sched_test}
    D_k + \min\limits_{1 \leq i \leq k} (D_i - D_i) \geq B_k + B_k' + \sum_{i=1}^{k} \left(\frac{C_i}{T_i}\right) (T_i - D_i + D_k),
\end{equation}
where $B_k = \sum_{i=1}^{k} C_i$ and $B_k' = \max\limits_{1 \leq i \leq k} \left(\max (0, S_i - C_i)\right)$ then the task set is schedulable under preemptive EDF.
\end{theorem}

\medskip

The key analysis in~\cite[Theorem 11]{DBLP:journals/tc/WangLZLT25}
extends the analysis by Devi~\cite{DBLP:conf/ecrts/Devi03}, which was
disproved by G\"unzel and Chen~\cite{DBLP:journals/rts/GunzelC20} in
2020. Devi's analysis for EDF scheduling extends the concept of
suspension as blocking proposed by
Liu~\cite[pp. 164–165]{Liu:2000:RS:518501} for fixed-priority
scheduling. Although Chen~et~al.~\cite{ChenHuangNelissen} prove the
correctness of the analysis proposed by Liu~\cite{Liu:2000:RS:518501},
they also note that, ``\emph{Even though the authors in this paper are
  able to provide a proof to support the correctness, the authors are
  not able to provide any rationale behind this method which treats
  suspension time as blocking time}''.  It is also noted in the review
paper by Chen et al. \cite{suspension-review-jj} that, ``\emph{It is in
  fact not clear why suspension induces blocking}.''

\section{Counterexamples}

\subsection{Counter Example for Theorem 11 by Wang~et~al.~\cite{DBLP:journals/tc/WangLZLT25}}
\label{sec:counterexample-theorem11}

\noindent We consider the following periodic task set $\mathbb{T} = \{\tau_1, \tau_2\}$: 

\begin{itemize}
    \item $\tau_1 = (C_1 = 2, S_1 = 6, D_1 = 8, T_1 = 12)$
    \item $\tau_2 = (C_2 = 2, S_2 = 10, D_2 = 18, T_2 = 18)$ 
\end{itemize}

\noindent According to Definition~\ref{def:necessary_cond}, the pre-condition holds for $\mathbb{T}$: 
\begin{align*}
    \frac{S_1}{T_1} + \frac{C_1}{T_1} + \frac{C_2}{T_2} = \frac{6}{12} + \frac{2}{12} + \frac{2}{18} = \frac{7}{9} \leq 1 \\ 
     \frac{S_2}{T_2} + \frac{C_1}{T_1} + \frac{C_2}{T_2} = \frac{10}{18} + \frac{2}{12} + \frac{2}{18} = \frac{5}{6} \leq 1 \\ 
\end{align*}
Since the pre-condition holds we evaluate the inequalities in Equation~\eqref{eq:sched_test} for both tasks next. 

\textbf{For task $\tau_1$:} 
\begin{align*}
    8 &= D_1 + \min\limits_{1 \leq i \leq 1} (D_i - D_i) \\
    &\geq B_1 + B_1' + \sum_{i=1}^{1} \left(\frac{C_i}{T_i}\right) (T_i - D_i + D_1) \\
    &=  2 + 4 + \left(\frac{2}{12}\right) (12 - 8 + 8) \\
    & = 8 
\end{align*}

\textbf{For task $\tau_2$:} 
\begin{align*}
    \frac{54}{3} &=  D_2 + \min\limits_{1 \leq i \leq 2} (D_i - D_i)\\
    &\geq B_2 + B_2' + \sum_{i=1}^{2} \left(\frac{C_i}{T_i}\right) (T_i - D_i + D_2) \\
    &= 4 + 8 + \left(\frac{2}{12}\right) (12 - 8 + 18) + \left(\frac{2}{18}\right) (18 - 18 + 18) \\
    &= \frac{53}{3}\\ 
\end{align*}

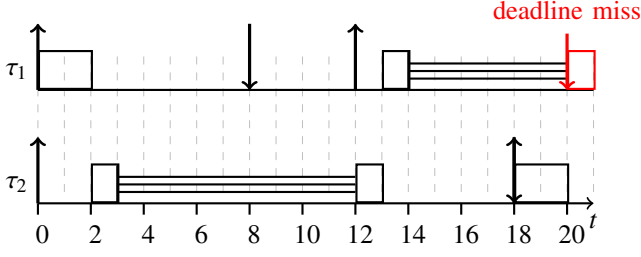
\begin{figure}[t]
    \centering
    \begin{tikzpicture}[x=\ux,y=\uy,auto, thick]
    \draw[->] (0,0) -- coordinate (xaxis) (21,0) node[anchor=north]{$t$};
    
    \foreach \x in {0,2,...,21}{
		\draw[-,below](\x,0) -- (\x,-0.3) node[] {\pgfmathtruncatemacro\yi{\x} \yi};
	}
	\foreach \x in {0,1,...,21}{
         \draw[-,very thin,lightgray, dashed](\x,0.3) -- (\x,4);
	}	
	\foreach \y in {3.0}{
		\draw[] (0,\y) -- (21,\y);
	}

	\begin{scope}[shift={(0,3)}]
		\node[anchor=east] at (0, 0.5) {$\tau_1$};
                \draw[->, very thick] (0,0) -- (0,1.75);
                \draw[->, very thick] (12,0) -- (12,1.75);
                \draw[<-, very thick] (8,0) -- (8,1.75);
                \node[task7, minimum width=2*\ux, anchor=south west] at (0, 0){};
                \node[task7, minimum width=\ux, anchor=south west] at (13, 0){};
                
                \node[task7, minimum width=\ux, anchor=south west, draw=red] at (20, 0){};

        		\draw[] (14,0) -- (14,1);
        		\draw[] (20,0) -- (20,1);
                
                \foreach \y in {0.3,0.5,0.7}{ 
        			\draw[] (14,\y) -- (20,\y);
                }
                \draw[<-,very thick, red] (20,0) -- (20,1.49);
                \node[anchor=south,red] at (20, 1.6) {deadline miss};
				
	\end{scope}

	\begin{scope}[shift={(0,0)}]
		\node[anchor=east] at (0, 0.5) {$\tau_2$};
        \draw[->, very thick] (0,0) -- (0,1.75);
        \draw[->, very thick] (18,0) -- (18,1.75);
        \draw[<-, very thick] (18,0) -- (18,1.75);
        \node[task7, minimum width=\ux, anchor=south west] at (2, 0){};
        \node[task7, minimum width=\ux, anchor=south west] at (12, 0){};
        \node[task7, minimum width=2*\ux, anchor=south west] at (18, 0){};

        \draw[] (3,0) -- (3,1);
        
        \foreach \y in {0.3,0.5,0.7}{ 
        			\draw[] (3,\y) -- (12,\y);
        }
	\end{scope}
      \end{tikzpicture}
      \caption{A concrete schedule for the task set in Section~\ref{sec:counterexample-theorem11}.
               Releases are denoted by upward arrows, and deadlines are denoted by downward arrows.
               Execution segments are denoted by rectangles, and suspension segments are denoted by three horizontal lines.}
          \label{fig:counterexample1}
\end{figure}

Since both inequalities hold, the test in Theorem~\ref{th:sched_test} concludes that the task set is schedulable under preemptive EDF. 
However, Figure~\ref{fig:counterexample1} depicts the first two job releases of both tasks. 
Due to the suspension of $\tau_2$ between $[2,12)$, $\tau_1$ becomes blocked by $\tau_2$ between $[12,13)$. 
This blocking delays the starting time of $\tau_1$'s second job, and hence, causes a deadline miss after $\tau_1$ becomes ready and suspends itself. 
Thus, the test is not sufficient to show a given task set is schedulable by EDF-Like scheduling. 

\subsection{Counter Example for Definition~\ref{def:necessary_cond}}

The necessary condition by Wang et al.~\cite{DBLP:journals/tc/WangLZLT25}, which is restated in Definition~\ref{def:necessary_cond}, indicates that
response time analysis methods are only applicable to systems that are
not overloaded. 
We disprove the necessary condition using the following task set $\mathbb{T} = \{\tau_1, \tau_2\}$:

\begin{itemize}
    \item $\tau_1 = (C_1 = 1, S_1 = 3, D_1 = 4, T_1 = 4)$
    \item $\tau_2 = (C_2 = 1, S_2 = 0, D_2 = 8, T_2 = 8)$ 
\end{itemize}

The necessary condition in Definition~\ref{def:necessary_cond} does not hold for $\mathbb{T} = \{\tau_1, \tau_2\}$ as $\tau_1$ violates it:
\begin{align*}
    \frac{S_1}{T_1} + \frac{C_1}{T_1} + \frac{C_2}{T_2} = \frac{3}{4} + \frac{1}{4} + \frac{1}{8} = \frac{9}{8} \nleq 1\ 
\end{align*}
Thus, Definition~\ref{def:necessary_cond} leads to a conclusion that
$\mathbb{T}$ is not feasible (i.e., some job misses its deadline)
under preemptive EDF.

We show that this task set is indeed feasibly scheduled under
preemptive EDF in both periodic and sporadic settings, invalidating
the necessary condition stated in Definition~\ref{def:necessary_cond}
for preemptive EDF.

\paragraph*{$\mathbb{T}$ is a periodic task set} 
When both $\tau_1$ and $\tau_2$ are periodic tasks and release their
first jobs at time $0$, the worst-case response time of $\tau_1$ is
$4$ and the worst-case response time of $\tau_2$ is $2$, as the jobs
of $\tau_2$ are executed whenever $\tau_1$ suspends itself or finishes
its execution. Therefore, no job misses its deadline under preemptive EDF.

\paragraph*{$\mathbb{T}$ is a sporadic task set} We apply the sufficient
schedulability test for EDF scheduling by Günzel et
al.~\cite[Theorem 12]{DBLP:conf/rtss/GunzelBCC22} with $b_1 = b_2 = 0$ and
$\widetilde{R}_1 = 4$, $\widetilde{R}_2 = 3$.  We obtain the following
response-time upper bounds for $\mathbb{T} = \{\tau_1, \tau_2\}$:

\textbf{Task $\tau_1$ (i.e., $k=1$):} 
\begin{align*}
    &\widetilde{R}_1(b_1) = \sum_{i\neq k} \max\left(\left\lceil \frac{\min(D_k - C_i, D_k - D_i) + \widetilde{R}_i - b_1}{T_i}\right\rceil, 0 \right) \\ &\cdot C_i + \left\lceil\frac{D_1 - b_1}{T_1}\right\rceil (C_1 + S_1) + b_1 \\ 
    & = \max\left(\left\lceil \frac{\min(4 - 1, 4 - 8) + 3 - 0}{8}\right\rceil, 0 \right) \cdot 1 
    + \left\lceil\frac{4 - 0}{4}\right\rceil (1 + 3) \\ 
    & = 4
\end{align*}

\textbf{Task $\tau_2$ (i.e., $k=2$):} 
\begin{align*}
    &\widetilde{R}_2(b_2) = \sum_{i\neq k} \max\left(\left\lceil \frac{\min(D_k - C_i, D_k - D_i) + \widetilde{R}_i - b_2}{T_i}\right\rceil, 0 \right) \\ &\cdot C_i + \left\lceil\frac{D_2 - b_2}{T_2}\right\rceil (C_2 + S_2) + b_2 \\ 
    & = \max\left(\left\lceil \frac{\min(8 - 1, 8 - 4) + 4 - 0}{4}\right\rceil, 0 \right) \cdot 1 
    + \left\lceil\frac{8 - 0}{8}\right\rceil (1 + 0) \\ 
    & = 3
\end{align*}

Since both $\widetilde{R}_1(b_1) =4 \leq D_1$ and $\widetilde{R}_2(b_2) =3 \leq D_2$, the task set $\mathbb{T}$ is schedulable by preemptive EDF scheduling. 

\section{Remark}
We note that in Wang~et~al.~\cite[Section~IV]{DBLP:journals/tc/WangLZLT25} also provide a jitter-based response time
analysis (called JRTA), extended from
Aromolo~et~al.~\cite{DBLP:conf/ecrts/AromoloBN22}. The counterexample
in Section~\ref{sec:counterexample-theorem11} of this paper does not invalidate the
JRTA results in~\cite{DBLP:journals/tc/WangLZLT25}.

\section*{Acknowledgment}

This result is part of a project (PropRT) that has received funding from the European Research Council (ERC) under the European Union’s Horizon 2020 research and innovation programme (grant agreement No. 865170). This work has been funded by the Deutsche Forschungsgemeinschaft (DFG, German Research Foundation) -- Project No. 569077889 (PEACH)
and by the National Aeronautics and Space Administration (NASA) award 80NSSC26K0764.

While the process of finding the counterexample of Theorem~11~in~\cite{DBLP:journals/tc/WangLZLT25}  was supported by the
use of AI, namely Anthropic Claude (Model Opus 5.5), the
counterexample presented in Section~\ref{sec:counterexample-theorem11} was hand-crafted for
readability.

\bibliographystyle{abbrv}
\bibliography{real-time}

\end{document}